



\font\twelverm=cmr10 scaled 1200    \font\twelvei=cmmi10 scaled 1200
\font\twelvesy=cmsy10 scaled 1200   \font\twelveex=cmex10 scaled 1200
\font\twelvebf=cmbx10 scaled 1200   \font\twelvesl=cmsl10 scaled 1200
\font\twelvett=cmtt10 scaled 1200   \font\twelveit=cmti10 scaled 1200

\skewchar\twelvei='177   \skewchar\twelvesy='60


\def\twelvepoint{\normalbaselineskip=12.4pt
  \abovedisplayskip 12.4pt plus 3pt minus 9pt
  \belowdisplayskip 12.4pt plus 3pt minus 9pt
  \abovedisplayshortskip 0pt plus 3pt
  \belowdisplayshortskip 7.2pt plus 3pt minus 4pt
  \smallskipamount=3.6pt plus1.2pt minus1.2pt
  \medskipamount=7.2pt plus2.4pt minus2.4pt
  \bigskipamount=14.4pt plus4.8pt minus4.8pt
  \def\rm{\fam0\twelverm}          \def\it{\fam\itfam\twelveit}%
  \def\sl{\fam\slfam\twelvesl}     \def\bf{\fam\bffam\twelvebf}%
  \def\mit{\fam 1}                 \def\cal{\fam 2}%
  \def\tt{\twelvett}
  \textfont0=\twelverm   \scriptfont0=\tenrm   \scriptscriptfont0=\sevenrm
  \textfont1=\twelvei    \scriptfont1=\teni    \scriptscriptfont1=\seveni
  \textfont2=\twelvesy   \scriptfont2=\tensy   \scriptscriptfont2=\sevensy
  \textfont3=\twelveex   \scriptfont3=\twelveex  \scriptscriptfont3=\twelveex
  \textfont\itfam=\twelveit
  \textfont\slfam=\twelvesl
  \textfont\bffam=\twelvebf \scriptfont\bffam=\tenbf
  \scriptscriptfont\bffam=\sevenbf
  \normalbaselines\rm}



\def\beginlinemode{\endmode
  \begingroup\parskip=0pt \obeylines\def\\{\par}\def\endmode{\par\endgroup}}
\def\beginparmode{\endmode
  \begingroup \def\endmode{\par\endgroup}}
\let\endmode=\par
{\obeylines\gdef\
{}}
\def\singlespace{\baselineskip=\normalbaselineskip}

\def\oneandahalfspace{\baselineskip=\normalbaselineskip
  \multiply\baselineskip by 3 \divide\baselineskip by 2}
\def\doublespace{\baselineskip=\normalbaselineskip \multiply\baselineskip by 2}

\newcount\firstpageno
\firstpageno=2
\footline={\ifnum\pageno<\firstpageno{\hfil}%
\else{\hfil\twelverm\folio\hfil}\fi}
\let\rawfootnote=\footnote              
\def\footnote#1#2{{\rm\singlespace\parindent=0pt\rawfootnote{#1}{#2}}}
\def\raggedcenter{\leftskip=4em plus 12em \rightskip=\leftskip
  \parindent=0pt \parfillskip=0pt \spaceskip=.3333em \xspaceskip=.5em
  \pretolerance=9999 \tolerance=9999
  \hyphenpenalty=9999 \exhyphenpenalty=9999 }
\def\dateline{\rightline{\ifcase\month\or
  January\or February\or March\or April\or May\or June\or
  July\or August\or September\or October\or November\or December\fi
  \space\number\year}}
\def\received{\vskip 3pt plus 0.2fill
 \centerline{\sl (Received\space\ifcase\month\or
  January\or February\or March\or April\or May\or June\or
  July\or August\or September\or October\or November\or December\fi
  \qquad, \number\year)}}


\hsize=6.5truein
\vsize=8.9truein
\parskip=\medskipamount
\twelvepoint            
\doublespace            
\overfullrule=0pt       



\def\title                      
  {\null\vskip 3pt plus 0.2fill
   \beginlinemode \doublespace \raggedcenter \bf}

\def\author                     
  {\vskip 3pt plus 0.2fill \beginlinemode
   \singlespace \raggedcenter}

\def\affil                      
  {\vskip 3pt plus 0.1fill \beginlinemode
   \oneandahalfspace \raggedcenter \sl}

\def\abstract                   
  {\vskip 3pt plus 0.3fill \beginparmode
   \doublespace \narrower ABSTRACT: }

\def\endtitlepage               
  {\endpage                     
   \body}

\def\body                       
  {\beginparmode}               

\def\head#1{                    
  \filbreak\vskip 0.5truein     
  {\immediate\write16{#1}
   \raggedcenter \uppercase{#1}\par}
   \nobreak\vskip 0.25truein\nobreak}

\def\subhead#1{                 
  \vskip 0.25truein             
  {\raggedcenter #1 \par}
   \nobreak\vskip 0.25truein\nobreak}

\def\refto#1{$|{#1}$}           

\def\references                 
  {\subhead{References}         
   \beginparmode
   \frenchspacing \parindent=0pt \leftskip=1truecm
   \parskip=8pt plus 3pt \everypar{\hangindent=\parindent}}

\gdef\refis#1{\indent\hbox to 0pt{\hss#1.~}}    

\gdef\journal#1, #2, #3, 1#4#5#6{               
    {\sl #1~}{\bf #2}, #3, (1#4#5#6)}           

\def\refstylenp{                
  \gdef\refto##1{ [##1]}                                
  \gdef\refis##1{\indent\hbox to 0pt{\hss##1)~}}        
  \gdef\journal##1, ##2, ##3, ##4 {                     
     {\sl ##1~}{\bf ##2~}(##3) ##4 }}

\def\refstyleprnp{              
  \gdef\refto##1{ [##1]}                                
  \gdef\refis##1{\indent\hbox to 0pt{\hss##1)~}}        
  \gdef\journal##1, ##2, ##3, 1##4##5##6{               
    {\sl ##1~}{\bf ##2~}(1##4##5##6) ##3}}

\def\endreferences{\body}

\def\figurecaptions             
  { \beginparmode
   \subhead{Figure Captions}
}

\def\endpage                    
  {\vfill\eject}

\def\endpaper                   
  {\endmode\vfill\supereject}
\def\endjnl
  {\endpaper}


\def\ref#1{Ref. #1}                     
\def\Ref#1{Ref. #1}                     

\def\frac#1#2{{\textstyle{#1 \over #2}}}
\def\half{{\textstyle{ 1\over 2}}}

\def\sla{\raise.15ex\hbox{$/$}\kern-.57em}
\def\leaderfill{\leaders\hbox to 1em{\hss.\hss}\hfill}
\def\twiddle{\lower.9ex\rlap{$\kern-.1em\scriptstyle\sim$}}
\def\bigtwiddle{\lower1.ex\rlap{$\sim$}}
\def\gtwid{\mathrel{\raise.3ex\hbox{$>$\kern-.75em\lower1ex\hbox{$\sim$}}}}
\def\ltwid{\mathrel{\raise.3ex\hbox{$<$\kern-.75em\lower1ex\hbox{$\sim$}}}}
\def\square{\kern1pt\vbox{\hrule height 1.2pt\hbox{\vrule width 1.2pt\hskip 3pt
   \vbox{\vskip 6pt}\hskip 3pt\vrule width 0.6pt}\hrule height 0.6pt}\kern1pt}

\def\m@th{\mathsurround=0pt }
\def\leftrightarrowfill{$\m@th \mathord\leftarrow \mkern-6mu
 \cleaders\hbox{$\mkern-2mu \mathord- \mkern-2mu$}\hfill
 \mkern-6mu \mathord\rightarrow$}
\def\overleftrightarrow#1{\vbox{\ialign{##\crcr
     \leftrightarrowfill\crcr\noalign{\kern-1pt\nointerlineskip}
     $\hfil\displaystyle{#1}\hfil$\crcr}}}


\font\titlefont=cmr10 scaled\magstep3

\def\martinstyletitle                      
  {\null\vskip 3pt plus 0.2fill
   \beginlinemode \doublespace \raggedcenter \titlefont}

\font\twelvesc=cmcsc10 scaled 1200

\def\author                     
  {\vskip 3pt plus 0.2fill \beginlinemode
   \singlespace \raggedcenter\twelvesc}


\def\heading                            
  {\vskip 0.5truein plus 0.1truein      
   \beginparmode \def\\{\par} \parskip=0pt \singlespace \raggedcenter}

\def\subheading                         
  {\vskip 0.25truein plus 0.1truein     
   \beginlinemode \singlespace \parskip=0pt \def\\{\par}\raggedcenter}

\def\tag#1$${\eqno(#1)$$}

\def\align#1$${\eqalign{#1}$$}

\def\aligntag#1$${\gdef\tag##1\\{&(##1)\cr}\eqalignno{#1\\}$$
  \gdef\tag##1$${\eqno(##1)$$}}

\def\endaligntag{}

\def\overset #1\to#2{{\mathop{#2}\limits^{#1}}}
\def\underset#1\to#2{{\let\next=#1\mathpalette\undersetpalette#2}}
\def\undersetpalette#1#2{\vtop{\baselineskip0pt
\ialign{$\mathsurround=0pt #1\hfil##\hfil$\crcr#2\crcr\next\crcr}}}


\def\ref#1{Ref.~#1}                     
\def\Ref#1{Ref.~#1}                     
\def\[#1]{[\cite{#1}]}
\def\cite#1{{#1}}
\def\(#1){(\call{#1})}
\def\call#1{{#1}}
\def\taghead#1{}
\def\frac#1#2{{#1 \over #2}}
\def\half{{\frac 12}}

\def\fourth{{\frac 14}}
\def\12{{1\over2}}

\def\sla{\raise.15ex\hbox{$/$}\kern-.57em}
\def\leaderfill{\leaders\hbox to 1em{\hss.\hss}\hfill}
\def\twiddle{\lower.9ex\rlap{$\kern-.1em\scriptstyle\sim$}}
\def\bigtwiddle{\lower1.ex\rlap{$\sim$}}
\def\gtwid{\mathrel{\raise.3ex\hbox{$>$\kern-.75em\lower1ex\hbox{$\sim$}}}}
\def\ltwid{\mathrel{\raise.3ex\hbox{$<$\kern-.75em\lower1ex\hbox{$\sim$}}}}
\def\square{\kern1pt\vbox{\hrule height 1.2pt\hbox{\vrule width 1.2pt\hskip 3pt
   \vbox{\vskip 6pt}\hskip 3pt\vrule width 0.6pt}\hrule height 0.6pt}\kern1pt}
\def\tdot#1{\mathord{\mathop{#1}\limits^{\kern2pt\ldots}}}

\def\pmb#1{\setbox0=\hbox{#1}%
  \kern-.025em\copy0\kern-\wd0
  \kern  .05em\copy0\kern-\wd0
  \kern-.025em\raise.0433em\box0 }

\catcode`@=11
\newcount\tagnumber\tagnumber=0

\immediate\newwrite\eqnfile
\newif\if@qnfile\@qnfilefalse
\def\write@qn#1{}
\def\writenew@qn#1{}
\def\w@rnwrite#1{\write@qn{#1}\message{#1}}
\def\@rrwrite#1{\write@qn{#1}\errmessage{#1}}

\def\taghead#1{\gdef\t@ghead{#1}\global\tagnumber=0}
\def\t@ghead{}

\expandafter\def\csname @qnnum-3\endcsname
  {{\t@ghead\advance\tagnumber by -3\relax\number\tagnumber}}
\expandafter\def\csname @qnnum-2\endcsname
  {{\t@ghead\advance\tagnumber by -2\relax\number\tagnumber}}
\expandafter\def\csname @qnnum-1\endcsname
  {{\t@ghead\advance\tagnumber by -1\relax\number\tagnumber}}
\expandafter\def\csname @qnnum0\endcsname
  {\t@ghead\number\tagnumber}
\expandafter\def\csname @qnnum+1\endcsname
  {{\t@ghead\advance\tagnumber by 1\relax\number\tagnumber}}
\expandafter\def\csname @qnnum+2\endcsname
  {{\t@ghead\advance\tagnumber by 2\relax\number\tagnumber}}
\expandafter\def\csname @qnnum+3\endcsname
  {{\t@ghead\advance\tagnumber by 3\relax\number\tagnumber}}

\def\equationfile{%
  \@qnfiletrue\immediate\openout\eqnfile=\jobname.eqn%
  \def\write@qn##1{\if@qnfile\immediate\write\eqnfile{##1}\fi}
  \def\writenew@qn##1{\if@qnfile\immediate\write\eqnfile
    {\noexpand\tag{##1} = (\t@ghead\number\tagnumber)}\fi}
}

\def\callall#1{\xdef#1##1{#1{\noexpand\call{##1}}}}
\def\call#1{\each@rg\callr@nge{#1}}

\def\each@rg#1#2{{\let\thecsname=#1\expandafter\first@rg#2,\end,}}
\def\first@rg#1,{\thecsname{#1}\apply@rg}
\def\apply@rg#1,{\ifx\end#1\let\next=\relax%
\else,\thecsname{#1}\let\next=\apply@rg\fi\next}

\def\callr@nge#1{\calldor@nge#1-\end-}
\def\callr@ngeat#1\end-{#1}
\def\calldor@nge#1-#2-{\ifx\end#2\@qneatspace#1 %
  \else\calll@@p{#1}{#2}\callr@ngeat\fi}
\def\calll@@p#1#2{\ifnum#1>#2{\@rrwrite{Equation range #1-#2\space is bad.}
\errhelp{If you call a series of equations by the notation M-N, then M and
N must be integers, and N must be greater than or equal to M.}}\else %
{\count0=#1\count1=#2\advance\count1 by1\relax\expandafter\@qncall\the\count0,%
  \loop\advance\count0 by1\relax%
    \ifnum\count0<\count1,\expandafter\@qncall\the\count0,%
  \repeat}\fi}

\def\@qneatspace#1#2 {\@qncall#1#2,}
\def\@qncall#1,{\ifunc@lled{#1}{\def\next{#1}\ifx\next\empty\else
  \w@rnwrite{Equation number \noexpand\(>>#1<<) has not been defined yet.}
  >>#1<<\fi}\else\csname @qnnum#1\endcsname\fi}

\let\eqnono=\eqno
\def\eqno(#1){\tag#1}
\def\tag#1$${\eqnono(\displayt@g#1 )$$}

\def\aligntag#1\endaligntag
  $${\gdef\tag##1\\{&(##1 )\cr}\eqalignno{#1\\}$$
  \gdef\tag##1$${\eqnono(\displayt@g##1 )$$}}

\def\eqalignno#1{\displ@y \tabskip\centering
  \halign to\displaywidth{\hfil$\displaystyle{##}$\tabskip\z@skip
    &$\displaystyle{{}##}$\hfil\tabskip\centering
    &\llap{$\displayt@gpar##$}\tabskip\z@skip\crcr
    #1\crcr}}

\def\displayt@gpar(#1){(\displayt@g#1 )}

\def\displayt@g#1 {\rm\ifunc@lled{#1}\global\advance\tagnumber by1
        {\def\next{#1}\ifx\next\empty\else\expandafter
        \xdef\csname @qnnum#1\endcsname{\t@ghead\number\tagnumber}\fi}%
  \writenew@qn{#1}\t@ghead\number\tagnumber\else
        {\edef\next{\t@ghead\number\tagnumber}%
        \expandafter\ifx\csname @qnnum#1\endcsname\next\else
        \w@rnwrite{Equation \noexpand\tag{#1} is a duplicate number.}\fi}%
  \csname @qnnum#1\endcsname\fi}

\def\ifunc@lled#1{\expandafter\ifx\csname @qnnum#1\endcsname\relax}

\let\@qnend=\end\gdef\end{\if@qnfile
\immediate\write16{Equation numbers written on []\jobname.EQN.}\fi\@qnend}

\catcode`@=12

\catcode`@=11
\newcount\r@fcount \r@fcount=0
\newcount\r@fcurr
\immediate\newwrite\reffile
\newif\ifr@ffile\r@ffilefalse
\def\w@rnwrite#1{\ifr@ffile\immediate\write\reffile{#1}\fi\message{#1}}

\def\writer@f#1>>{}
\def\referencefile{
  \r@ffiletrue\immediate\openout\reffile=\jobname.ref%
  \def\writer@f##1>>{\ifr@ffile\immediate\write\reffile%
    {\noexpand\refis{##1} = \csname r@fnum##1\endcsname = %
     \expandafter\expandafter\expandafter\strip@t\expandafter%
     \meaning\csname r@ftext\csname r@fnum##1\endcsname\endcsname}\fi}%
  \def\strip@t##1>>{}}

\def\citeall#1{\xdef#1##1{#1{\noexpand\cite{##1}}}}
\def\cite#1{\each@rg\citer@nge{#1}} 

\def\each@rg#1#2{{\let\thecsname=#1\expandafter\first@rg#2,\end,}}
\def\first@rg#1,{\thecsname{#1}\apply@rg} 
\def\apply@rg#1,{\ifx\end#1\let\next=\relax
\else,\thecsname{#1}\let\next=\apply@rg\fi\next}

\def\citer@nge#1{\citedor@nge#1-\end-} 
\def\citer@ngeat#1\end-{#1}
\def\citedor@nge#1-#2-{\ifx\end#2\r@featspace#1 
  \else\citel@@p{#1}{#2}\citer@ngeat\fi} 
\def\citel@@p#1#2{\ifnum#1>#2{\errmessage{Reference range #1-#2\space is bad.}%
    \errhelp{If you cite a series of references by the notation M-N, then M and
    N must be integers, and N must be greater than or equal to M.}}\else%
 {\count0=#1\count1=#2\advance\count1 by1\relax\expandafter\r@fcite\the\count0,
  \loop\advance\count0 by1\relax
    \ifnum\count0<\count1,\expandafter\r@fcite\the\count0,%
  \repeat}\fi}

\def\r@featspace#1#2 {\r@fcite#1#2,} 
\def\r@fcite#1,{\ifuncit@d{#1}
    \newr@f{#1}%
    \expandafter\gdef\csname r@ftext\number\r@fcount\endcsname%
                     {\message{Reference #1 to be supplied.}%
                      \writer@f#1>>#1 to be supplied.\par}%
 \fi%
 \csname r@fnum#1\endcsname}
\def\ifuncit@d#1{\expandafter\ifx\csname r@fnum#1\endcsname\relax}%
\def\newr@f#1{\global\advance\r@fcount by1%
    \expandafter\xdef\csname r@fnum#1\endcsname{\number\r@fcount}}

\let\r@fis=\refis   
\def\refis#1#2#3\par{\ifuncit@d{#1}
   \newr@f{#1}%
   \w@rnwrite{Reference #1=\number\r@fcount\space is not cited up to now.}\fi%
  \expandafter\gdef\csname r@ftext\csname r@fnum#1\endcsname\endcsname%
  {\writer@f#1>>#2#3\par}}

\def\ignoreuncited{
   \def\refis##1##2##3\par{\ifuncit@d{##1}%
    \else\expandafter\gdef\csname r@ftext\csname r@fnum##1\endcsname\endcsname%
     {\writer@f##1>>##2##3\par}\fi}}

\def\r@ferr{\endreferences\errmessage{I was expecting to see
\noexpand\endreferences before now;  I have inserted it here.}}
\let\r@ferences=\references
\def\references{\r@ferences\def\endmode{\r@ferr\par\endgroup}}

\let\endr@ferences=\endreferences
\def\endreferences{\r@fcurr=0
  {\loop\ifnum\r@fcurr<\r@fcount
    \advance\r@fcurr by 1\relax\expandafter\r@fis\expandafter{\number\r@fcurr}%
    \csname r@ftext\number\r@fcurr\endcsname%
  \repeat}\gdef\r@ferr{}\endr@ferences}


\let\r@fend=\endpaper\gdef\endpaper{\ifr@ffile
\immediate\write16{Cross References written on []\jobname.REF.}\fi\r@fend}

\catcode`@=12

\citeall\refto  
\citeall\ref  %
\citeall\Ref  %

\vglue 0. truein
\title
{
Detailed Stability Analysis of Electroweak Strings
}
\author
{Margaret James}
\affil
{
D.A.M.T.P., Silver Street,
University of Cambridge, Cambridge, CB39EW, U.K.
}
\smallskip
\author
{Leandros Perivolaropoulos}
\affil
{
Division of Theoretical Astrophysics,
Harvard-Smithsonian Center for Astrophysics,
60 Garden Street, Cambridge, MA 02138.
}
\smallskip
\author
{Tanmay Vachaspati}
\affil
{
Tufts Institute of Cosmology, Department of Physics and Astronomy,
Tufts University, Medford, MA 02155.
}

\abstract

We give a detailed stability analysis
of the Z-string in the standard
electroweak model. We identify the mode that determines the
stability of the string and numerically map the region of parameter
space where the string is stable. For $\sin^2 \theta_W
= 0.23$, we find that the strings are unstable for a
Higgs mass larger than 23GeV. Given the latest constraints
on the Higgs mass from LEP, this shows that, if the standard
electroweak model is realized in Nature, the existing vortex solutions
are unstable.

\beginsection{1. Introduction}

Recent studies have shown that the stability of topological defects may persist
when they are embedded in theories where they are not topologically stable.
Such ``embedded defects'' are exact solutions of the equations of motion in
the theories where they are embedded and can be dynamically stable.

A typical example of such a defect is the
semilocal string [1,2],
an embedding of the Nielsen-Olesen vortex in a ``semilocal'' model with
$SU(2)_{global}\times U(1)_{local}\rightarrow U(1)_{global}$
symmetry breaking. The semilocal string has been shown to be stable for a
finite parameter sector [3,4]
even though the vacuum manifold in the semilocal
model is $S^3$. In fact, since $\pi_1(S^3)=1$, the stability of the
semilocal string is dynamical rather than topological.
The crucial reason for this stability is that for a certain
parameter region, the
increase in gradient energy necessary for the string to decay
is more than the
corresponding decrease in
potential energy. Therefore, for that
parameter region, the decay would increase the total energy and is not
favored energetically.

The obvious generalization of the semilocal model is the electroweak model
in which the $SU(2)$ symmetry becomes gauged: $SU(2)_L \times U(1)_Y
\rightarrow U(1)_{em}$. It is possible to show that
the Nielsen-Olesen vortex may be embedded in the electroweak model
i.e. the electroweak equations of motion with a vortex-like ansatz
reduce to the Nielsen-Olesen equations [5,6].
Two such embedded vortex solutions - the so-called $\tau -$
and $Z-$ strings - are known [7]. Here we shall only consider the
$Z-$string as the $\tau -$string is expected to be unstable
for all values of the parameters [7].

The phenomenological successes of the standard electroweak
model [8] make the question of stability of the electroweak string
a very interesting and important one. A stable electroweak string would
for the first time in particle physics provide a macroscopic, stable
coherent state at low enough energies to be accessible to particle physics
experiments. It is this question of stability [9] that we
are addressing in detail in this paper. In particular, we construct a map
of the parameter space of the standard electroweak model showing the
range of parameters where the $Z-$string is stable and where it is
unstable. Here we give the details of the calculation as well as
the physical reasoning for the simplifications that occur in this
originally highly complicated problem. A brief exposition of the
main results we derive here may be found in Ref.[15].

The structure of the paper is the following: in the next section we give
a review of the electroweak string showing that the electroweak
equations of motion reduce to the Nielsen-Olesen equations for a
particular vortex ansatz. In section 3 we show how can the stability
problem of the electroweak string be reduced to the eigenvalue problem
of a single Schroedinger-like eigenvalue problem.
This is a non-trivial simplification since the initial system
of coupled perturbations involves twenty coupled degrees of freedom
which after tedious manipulations not only decouple but also reduce to
a single eigenvalue equation. Finally in section 4
we solve this eigenvalue problem and construct a map showing the
parameter sector corresponding to stability.

\beginsection{2. Review of Electroweak Strings}
\taghead{2.}

We consider static bosonic field configurations in the Weinberg-Salam model:
there is no time dependence and we choose a gauge where the zero components of
the gauge fields are set
to zero. The energy functional is given by
$$
E = \int d^3 x\left [
          \fourth G_{ij} ^a G_{ij} ^a + \fourth F_{Bij} F_{Bij}
          + (D_j \phi ) ^{\dag} (D_j \phi ) +
            \lambda (\phi ^{\dag} \phi - \eta ^2 /2 )^2
              \right ]
\eqno (2.1)
$$

The Weinberg
angle is given by tan $\theta_w={g'/g}$. The masses of the
W-boson, the Z-boson and Higgs boson are, respectively,
$$
M_W={1\over
2}g\eta ,\qquad\ M_Z={1\over 2}\alpha\eta ,\qquad
M_H=\sqrt{2\lambda}\eta\ ,
\eqno(2.2)
$$
where
$$
\alpha \equiv \sqrt{ g^2 + {g'} ^2 }
\eqno(2.3)
$$
The time-independent field equations are
$$
D_j F^a_{ij}=- {\textstyle {1\over2}}ig(\phi^{\dagger}
\ \tau^a D_i \phi - (D_i \phi)^{\dagger} \ \tau^a \phi)
\eqno(2.4)
$$
$$
\partial_jf_{ij}=- {\textstyle{1\over2}}ig'\left(\phi^{\dagger}D_i\phi-(D_i
\phi)^{\dagger}\phi\right)
\eqno(2.5)
$$
$$
D_iD_i\phi=
       2\lambda(\phi^{\dagger}\phi-{\textstyle{1\over 2}}\eta^2)\phi\ ,
\eqno(2.6)
$$
The symbols are in the standard notation
defined in Ref. 10.
In addition, we recall the usual mixing formula:
$$
Z^\mu \equiv cos\theta_W W^{\mu 3} - sin\theta_W B^\mu \  ,
\ \ \ \
A^\mu \equiv sin\theta_W W^{\mu 3} + cos\theta_W B^\mu \  ,
\eqno (2.7)
$$

The vortex solution extremising (2.1) is given by [5,6]:
$$
\eqalign {
W^{\mu 1} = 0 & = W^{\mu 2} = A^\mu , \ \ \
Z^\mu = [ A^\mu ] _{NO} = - {{v_{NO} (r)} \over r} {\hat e}_\theta
\cr
&
\phi = f_{NO} (r) e^{im \theta } \Phi \ , \ \ \
\Phi \equiv \pmatrix{0\cr 1\cr}
\cr
}
\eqno (2.8)
$$
where, the coordinates $r$ and $\theta$ are polar coordinates
in the $xy-$plane. The integer $m$ is the winding number of the
vortex and, here, we shall restrict ourselves to the case
$m=1$. The
subscript $NO$ on the functions $f$ and
$A^\mu$ means
that they are identical to the corresponding functions found
by Nielsen and Olesen [11] for the usual Abelian-Higgs string.
On substituting eq. \(2.8) into the equations of motion they reduce to
$$
f'' + {{f'} \over r} - \left ( 1- {\alpha \over 2} v \right ) ^2
                             {f \over {r^2}}
   - 2 \lambda \left ( f^2 - {{\eta ^2} \over 2} \right ) f = 0
\eqno (2.9)
$$
$$
v'' - {{v'} \over r} + \alpha \left ( 1 - {\alpha \over 2} v \right ) f^2 = 0
\eqno (2.10)
$$
where primes denote differentiation with respect to $r$
and the subscripts $NO$ have been dropped for convenience. These are solved
together with the
boundary conditions:
$$
f(0) = 0 = v(0), \ \ \  f(\infty ) = {\eta \over {\sqrt{2}}} , \ \ \
v(\infty ) = {2 \over \alpha}
\eqno (2.11)
$$
The string solutions resulting from these equations have been studied
previously by several authors in
a lot of detail. A sample of these papers may be found in the
collection of Ref. 12.

At this point, it is useful to note the symmetries of the string configuration.
Firstly it is axially symmetric i.e. it is invariant under the action of the
symmetry operator generated by the generalised angular momentum operator $$
K_z\ =\ L_z\ +\ S_z\ + \ I_z\ . \eqno(2.12) $$ $L_z$ and $S_z$ are the usual
orbital and spin pieces respectively of the spatial angular momentum operator.
Explicitly $${L_z}=-i{\partial\over \partial\theta}\ {\bf
1}\qquad\left({S_z}{\vec a}\right)_j=-i\epsilon_{3jk}{\vec a}_k\ {\bf 1}\
,\eqno(2.13)$$ where $\vec a$ is any vector field and ${\bf 1}$ is the $2\times
2$ unit matrix. . Note that $S_z$ annihilates the scalar Higgs field. $I_z$ is
composed of a $U(1)$ generator, $Y$ and an $SU(2)$ generator, $T^3$

$$ I_z= -\half( Y-T^3)\ . \eqno(2.14) $$
$Y$ and $T^3$ act on the Higgs field on the left. $Y$ annihilates the gauge
field and $T^3$ acts via a commutator bracket.

The configuration has two further symmetries. It is invariant under the
combination of reflection in the x-axis and complex conjugation. Also it is
invariant under the action of the global $U(2)$ gauge tranformation given
by $${\overline U}=\pmatrix{-1&0\cr 0&1\cr}\eqno(2.15)$$ Our eventual expansion
 of the
perturbations will be in Fourier modes but we shall point out some connections
between these and the eigenfunctions of the above commuting symmetry operators.


\beginsection{3. Stability of Electroweak Strings}
\taghead{3.}

 The vortex solution given in eq. \(2.8) is not topologically stable. This
means
that any field configuration can be continuously deformed to the vacuum. Hence
we are investigating the metastability of the vortex solution i.e.
whether it is a {\it local} maximum or minimum in configuration space. We
consider infinitesimal perturbations of the vortex configuration and ask if the
variation in the energy is positive or negative.

Let us write
$$
\phi = \pmatrix{\phi_1\cr \phi_{NO} + \phi_2\cr}
\eqno (3.1)
$$
$$
Z^\mu = Z_{NO} ^\mu + \delta Z^\mu
\eqno (3.2)
$$
$$
T^1 \equiv diag( - cos2\theta_W , 1 ),
\eqno (3.3)
$$
and,
$$
{\bf d} _j \equiv ( \partial _j {\bf 1} + i\half \alpha T^1 Z_j ) \  .
\eqno (3.4)
$$

The perturbations can depend on the $z-$coordinate and the $z-$components
of the vector fields can be non-zero also. However, since the vortex
solution has translational invariance along the $z-$direction, it is
easy to see that the $z-$dependence in the perturbations can be ignored
and the $z-$components of the gauge fields can be set to zero.
This follows from (2.1) where
the relevant $z-$dependent terms in the integrand are:
$$
          \half G_{i3} ^a G_{i3} ^a + \fourth F_{Bi3} F_{Bi3}
          + (D_3 \phi ) ^{\dag} (D_3 \phi )
\eqno (3.5)
$$
This contribution to the energy
is strictly non-negative and is minimized (that is, made to vanish)
by setting the $z-$components of the gauge fields
to zero and also considering the perturbations to be independent of the
$z-$coordinate. For this reason,
we shall drop all reference to the $z-$coordinate
in the calculations below and it will be understood that the energy is
actually the energy {\it per unit length} of the string.

Now we calculate the energy of the perturbed configuration
discarding terms of cubic and higher
order in the infinitesimal perturbations. We find,
$$
E = ( E_{NO} + \delta E_{NO} ) + E_1 + E_c + E_W
\eqno (3.6)
$$
where,
$E_{NO}$ is the energy of the Nielsen-Olesen string and $\delta E_{NO}$ is
the energy variation due to the perturbations $\phi_2$ and $\delta Z^\mu$.
The variation $E_1$ is due to the perturbation $\phi_1$ in the upper
component of the Higgs field:
$$
E_{1} = \int d^2 x \left [
       |{\bar d}_j \phi_1 |^2 + 2 \lambda ( f^2 - \eta^2 /2 ) | \phi_1 |^2
                         \right ] \  ,
\eqno (3.7)
$$
where,
$$
{\bar d} _j \equiv \partial _j - i{\alpha \over 2} cos(2\theta_W ) Z_j \  .
\eqno (3.8)
$$
The contribution from the $\phi$ and
${\vec W}^{\bar a}$ [13] interaction is:
$$
E_c = cos\theta_W \int d^2 x J_j ^{\bar a} W_j ^{\bar a}
\eqno (3.9)
$$
$$
J_j ^{\bar a} \equiv \half i \alpha \left [
                     \phi^{\dag} \tau^{\bar a} {\bf d}_j \phi
         - ( {\bf d}_j \phi ) ^{\dag} \tau^{\bar a} \phi \right ]
\eqno (3.10)
$$
and the energy in the ${\vec W}^{\bar a}$ and $\vec A$ bosons is
$$
\eqalign {
E_W \equiv
&
           \int d^2 x \biggl [
            \gamma {\vec W} ^1 \times {\vec W} ^2 \cdot
                                                 \vec \nabla \times \vec Z
+ \half |\vec \nabla \times {\vec W} ^1
                  + \gamma {\vec W} ^2 \times \vec Z | ^2
\cr &\cr
&+
\half |\vec \nabla \times {\vec W} ^2
                  + \gamma \vec Z \times {\vec W} ^1 | ^2
+ \fourth g^2 f^2 ( {\vec W } ^{\bar a} ) ^2
+ \half ( \vec \nabla \times \vec A )^2
                     \biggr ] \  .
\cr}
\eqno (3.11)
$$
where, $\gamma \equiv g cos\theta_W$. It may be remarked that
the $f$ and $\vec Z$ fields in eqs. (3.7)-(3.11)
are the unperturbed fields of the string since
we are only keeping up to quadratic terms in the infinitesimal
quantities.

Firstly we note that the perturbations of the fields that make up the string do
not couple to the other available perturbations. i.e. the perturbations in the
fields $f$ and $v$ only occur inside the variation $\delta E_{NO}$. We can
understand this as follows: the perturbation of the string solution has
${\overline U}=1$, where ${\overline U}$ is given by
eq. \(2.15)., whereas the other perturbations have
${\overline U}=-1$. Now,
since we know that the Nielsen-Olesen string with unit winding number is stable
to perturbations for any values of the parameters then necessarily, $\delta E
_{NO} \ge 0$ and the perturbations $\phi_2$ and $\delta Z^\mu$ cannot
destabilize the vortex. Then, we are justified in ignoring these perturbations
and setting $\delta E _{NO} = 0$. Also we note that the ${\vec A}$ boson only
appears in the last term of eq. \(3.11) and obviously makes a positive
contribution so we can set ${\vec A}$ to zero.


We now consider an expansion of the remaining perturbations in Fourier modes.
This gives,
$$
\phi _1 = \chi (r) e^{im\theta}
\eqno (3.12)
$$
for the $m^{th}$ mode where $m$ is any integer. For the gauge fields
we have,
$$
{\vec W}^1 = \left [
\left \{ {\bar f}_{1} (r) cos(n\theta ) + f_{1} sin(n\theta ) \right \}
{\hat e} _r +
{1 \over r} \left \{
           - {\bar h}_{1} sin(n\theta ) + h_{1} cos(n \theta ) \right \}
{\hat e} _\theta
\right ]
\eqno (3.13)
$$
$$
{\vec W}^2 = \left [
\left \{ - {\bar f}_{2} (r) sin(n\theta ) + f_{2} cos(n\theta ) \right \}
{\hat e} _r +
{1 \over r} \left \{
           {\bar h}_{2} cos(n\theta ) + h_{2} sin(n \theta ) \right \}
{\hat e} _\theta
\right ]
\eqno (3.14)
$$
for the $n^{th}$ mode where $n$ is a non-negative integer.
Inserting the expressions for the $m^{th}$ mode of $\phi_1$ and the
$n^{th}$ mode of the ${\vec W}^{\bar a}$ fields
in the energy functional gives:
$$
E_1 =
       2\pi \int dr \  r \left [
           {\chi '}^2 + \left \{
              {1 \over {r^2}}
            \left ( m + { \alpha \over 2} cos 2\theta_W v \right ) ^2 +
             2 \lambda \left ( f^2 - {{\eta ^2} \over 2} \right )
                     \right \} \chi ^2 \right ]
\eqno (3.15)
$$
$$
\eqalign{
E_c =  \delta_{\pm n,1-m} \pi \alpha cos\theta_W
          \int dr r
&
\biggl [ -(f\chi ' - \chi f') (f_2 \mp f_1 )
\cr
&
             - {f \over {r^2}} \chi \left \{
                \mp n + 2 - {\alpha \over 2} (1-cos2\theta_W ) v \right \}
               (h_2 \pm h_1 )
\biggr ]
\cr
}
\eqno (3.16)
$$
for $n \ne 0$. In the case when $n=0$, the expression for $E_c$ is equal to the
above expression multiplied by a factor of 2 and
with $f_1$ and $h_2$ set equal to zero. Finally,
$$
\eqalign{
E_W = \pi \int {{dr} \over r}
&
\biggl [
                   \gamma ( f_2 h_1 - f_1 h_2 ) v'
\cr
&
+
\half \left |- n f_1 + {h_1} ' - \gamma v f_2 \right | ^2 +
\half \left | n f_2 + {h_2} ' + \gamma v f_1 \right | ^2
\cr
&
+
\fourth g^2 f^2
[ r^2 ( {f_1} ^2 + {f_2} ^2 ) + {h_1}^2 + {h_2}^2 ] \biggr ] +
( f_{\bar a} \rightarrow {\bar f} _{\bar a} ,
  h_{\bar a} \rightarrow {\bar h} _{\bar a} ) \  .
\cr }
\eqno (3.17)
$$
for $n \ne 0$. In the case when $n=0$, $E_W$ is given by (3.17) multiplied
by a factor of 2 and with $f_1$ and $h_2$ set equal to zero.

First let us inspect $E_1$. Here the negative contributions can come
from the term proportional to $\chi ^2$. The coefficient of $\chi ^2$ is
composed
of two terms: the second term comes from  comes from the potential part and is
negative while the first term comes from the kinetic part and is always
positive and is smallest when $m=0$, at least in
the region near the center of the string where an instability is most
likely to develop. That is, the $m=0$ mode is the ``most dangerous'' mode.

Next we inspect $E_W$. Here the analysis is less obvious. Yet one can see
that the only term that can be negative is the first term that arises from
the term ${\vec W} ^1 \times {\vec W} ^2 \cdot \vec \nabla \times \vec Z$
in eq. (3.11). For this to contribute at the center of the string, the
vector fields must not vanish there. But the only mode that need not vanish
at the center is the $n=1$ mode; all other modes have to vanish at the
center if they are to be single-valued and finite. Therefore the only mode
that can give negative contributions at the center of the string is the
$n=1$ mode. Hence, we restrict ourselves to considering the
$m=0, n=1$ mode.

There is also a generally accepted idea which leads us to expect the least
stable mode to be the $m=0$, $n=1$ mode. As noted, the original configuration
has $K_z=0$ and ${\overline U}=1$ . The $m=0$ mode is also
$K_z=0$ with ${\overline U}=-1$. Turning our attention to the gauge fields
since
$$K_z\ e^{in\phi}\tau_\pm=(n\pm 1)\ e^{in\phi}\tau_\pm\ ,\eqno(3.18)$$ where
$\tau_\pm=\tau_1\pm i\tau_2$, we see that taking the combinations $f_1 - f_2$
 and
$h_1 + h_2$ we get $K_z=0$ and
${\overline U}=-1$. Intuitively we expect the least stable mode to have
$K_z=0$ since as $\bigl\vert K_z\bigr\vert$ increases one gets an increasing
 centrifugal barrier. The
other combinations $f_1 + f_2$ and $h_1 - h_2$ are a superposition of $K_z=\pm
2$ and we will see they decouple and drop out of our analysis.

Also note that
the barred and unbarred perturbations decouple. ( Under the combined operation
of reflection in the x-axis and complex
conjugation the barred and unbarred variables have
opposite sign.) Furthermore, the stability problem
in the barred variables is contained within the problem of the unbarred
variables (all we need to do is to set $\phi_1 = 0$). Therefore, it is
sufficient to consider only the unbarred functions.

We will now systematically simplify the expression for the energy variation.
After a lot of algebra, we obtain the first step:
\medskip

$$\eqalign{{\delta E\over 2\pi} = &
       \int dr \  r \biggl [
          \left( 1- {{(grf)^2} \over 2{P_+}} \right) {\chi '}^2 +
               M^2 \chi ^2 \biggr ]
\cr
&\ \cr
       &+ \int {dr \over 2r} \biggl [
           {A_{\pm}} {\xi_{\pm} '}^2 +
               S_{\pm} {\xi_{\pm} }^2 \biggr ]
\cr
&\ \cr
 &+ \alpha cos\theta_W \int dr \  r
         \biggl [{\{(1-\gamma v)\xi_+'+\gamma v'\xi_+\} \over P_+}(f\chi
'-f'\chi)-
           {f\chi \over {r^2}} (1-\alpha sin^2 \theta_W v ) \xi_+
              \biggr ]
\cr &\ \cr
& +T_+(F_-,\chi,\xi_+)+T_-(F_+,\xi_-)\cr}
\eqno (3.19)
$$
where,
$${T_+(F_-,\chi,\xi_+)=\int {dr\over 2r}\biggl [
   {\sqrt{P_+}} F_- +
   { {\{ (1-\gamma v) \xi_+ ' + \gamma v' \xi _+ \} } \over
       {\sqrt{P_+}}}
- \alpha cos\theta_W r^2 {{(f\chi ' - f' \chi )} \over {\sqrt{P_+}}}
                     \biggr ] ^2
}\eqno (3.20)
$$
$$T_-(F_+,\xi_-)=\int {dr \over 2r}\biggl [
          {\sqrt{P_-}} F_+ +
   { {\{ (1+\gamma v) \xi_- ' - \gamma v' \xi _- \} } \over
       {\sqrt{P_-}}}\biggr]^2
\eqno(3.21)
$$
$$P_{\pm} = (1 \mp \gamma v )^2 + {g^2 r^2 f^2\over 2}
\eqno (3.22)
$$
$$
F_{\pm} = {{f_2 \pm f_1} \over {2}}
\eqno (3.23)
$$
$$
\xi_{\pm} = {{h_2 \pm h_1} \over 2}
\eqno (3.24)
$$
$$
A_\pm (r) = {{g^2 r^2 f^2} \over {2P_\pm (r)}}
\eqno (3.25)
$$
\
$$
M^2 = {{\alpha ^2} \over {4 r^2}} cos^2 2\theta_W v^2 +
             2 \lambda \left ( f^2 - {{\eta ^2} \over 2} \right )
      - {{(grf')^2} \over 2{P_+}}  -
{1 \over r} {d \over {dr}} \left ( {{g^2 r^3 f f'} \over 2{P_+}} \right )
\eqno (3.26)
$$
\

$$
S_\pm (r) =  {g^2 f^2\over 2} - {{\gamma ^2 {v'}^2} \over {P_\pm (r)}} \pm
    r {{d \  } \over {dr}} \left [
            \gamma {{v'} \over r} {{(1 \mp \gamma v )} \over {P_\pm (r)}}
                          \right ] \  .
\eqno (3.27)
$$

As expected the problem in $\chi, F_-$ and $\xi_+$ has decoupled from the
problem in $F_+$ and $\xi_-$. Furthermore the only terms containing $F_-$ and
$F_+$ are $T_-$ and $T_+$ respectively and since these are whole squares
they can be set to zero. This then leaves us with a problem in  $\chi$
and $\xi_+$ and a problem in just $\xi_-$.

We first discuss the $\xi_-$ problem. We conjecture that the relevant
potential, $S_-$ is positive for any values of the parameters $\beta$
and $\cos\theta_W$ (where $\beta=8\lambda/g^2$). We motivate this by
considering
the asymptotic behaviour as $r\rightarrow 0$ and as $r\rightarrow \infty$: in
both these limits $S_-$ is always positive. Our conjecture is therefore
reasonable since the function is basically exponential, modulated by
polynomials. (We have checked it numerically for many pairs of parameters.)

Hence our analysis is reduced to considering perturbations in $\xi_+$ and
$\chi$ alone. We express the change in the energy in the form of an eigenvalue
problem:
$$
\delta E[ \chi , \xi_+ ] = 2\pi \int dr \  r
              ( \chi , \xi _+ ) {\bf O} \pmatrix{\chi\cr \xi_+\cr}
\eqno (3.28)
$$
where, {\bf O} is a $2\times 2$ matrix differential operator.

It is now useful to identify the form of the perturbations $\chi$
and $\xi_+$ that are pure gauge transformations of the string
configuration.
It is easy to see that perturbations of the form
$$
\delta\phi=ig\psi\phi_0,{\ {\delta W}_i=-iD_{0i}\psi} \  ,
\eqno (3.29)
$$
where $\psi$ is a real {\it L(SU(2))}
valued function and the $0$ subscript denotes the unperturbed
fields, is an infinitesimal gauge transformation of the
original
vortex solution. (In (3.29), $W_i$ represents
$\vec \tau \cdot \vec W _i$.) If we now require that these
purely gauge perturbations do not affect the
string configuration itself, then we can only have
$$
\psi=s(r)\pmatrix{0&ie^{-i\phi}\cr -ie^{i\phi}&0\cr}
\eqno (3.30)
$$
where $s(r)$ is any smooth function. This means that
perturbations given by
$$
\pmatrix{\chi\cr\xi_+}\ =\ s(r)\pmatrix{-gf\cr 2(1-\gamma v)\cr}
\eqno(3.31)
$$
are pure gauge perturbations that do not affect the string configuration.
Therefore, such perturbations cannot contribute to the energy variation
and must be annihilated
by ${\bf O}$. Then, in the two-dimensional
space of $(\chi , \xi_+ )$ perturbations, we can choose a basis in
which one direction is pure gauge and is given by (3.31) and the other
orthogonal direction is the direction of physical perturbations.
The physical mode is,
$$
\zeta\ =\ (1-\gamma v)\chi\ +\ {gf\over 2}\xi_+\quad .
\eqno(3.32)
$$

It was a good check on our algebra that on eliminating $\xi_+$ in terms of
$\zeta$ and $\chi$ in eq. (3.28)
the functional reduces to one depending only on
$\zeta$:
$$
\delta E[ \zeta ] = 2\pi \int dr \  r
              \zeta  {\overline O} \zeta
\eqno (3.33)
$$
where ${\overline O}$ is the differential operator
$${\overline O} = - {1 \over r} {d \over {dr}} \left (
             {r \over {P_+}} {d \over {dr}} \right ) + U(r)
\eqno (3.34)$$
and
$$ U(r) = {{{f'}^2} \over {P_+ f^2}} + {{2 S_+} \over {g^2 r^2 f^2}} +
          {1 \over r} {d \over {dr}} \biggl (
                     {{r f'} \over {P_+ f}} \biggr ) \  .\eqno (3.35)$$

To summarise, the question of stability reduces to asking if the operator
${\overline O}$ has negative eigenvalues in its spectrum. That is, whether the
eigenvalue $\omega$ of the Schrodinger equation,
$$ {\overline O} \zeta =\omega\zeta \  , \eqno (3.36) $$
can be negative. The eigenfunction $\zeta$ must also
satisfy the boundary conditions $\zeta (r=0) = 1$ and $\zeta \rightarrow c$
($c$ is some constant) as $r \rightarrow\infty$.

In this way we have reduced the stability analysis down to one Schrodinger
equation which we will solve numerically.


\beginsection{4. Numerical Analysis}
\taghead{4.}

We solve eq. \(3.36) together with the Nielson-Olesen eqs. (2.9) and (2.10).
However, it is convenient to work with rescaled dimensionless variables.
Hence, we define
$$
P \equiv {{\sqrt{2}} \over \eta} f , \ \ \ \
V \equiv {\alpha \over 2} v , \ \ \ \
R \equiv {{\alpha \eta} \over {2\sqrt{2}}} r \  .
\eqno (4.1)
$$
In terms of these dimensionless variables, the Nielsen-Olesen equations
\(2.9) and \(2.10) become,
$$
P'' + {{P'} \over R} - \left ( 1- V \right ) ^2
                             {P \over {R^2}}
   + \beta ( 1 - P^2 ) P = 0
\eqno (4.2)
$$
$$
V'' - {{V'} \over R} + 2 ( 1 - V ) P^2 = 0
\eqno (4.3)
$$
where primes now denote differentiation with respect to $R$. The functions
$P$ and $V$ also satisfy the boundary conditions:
$$
P(0) = 0 = V(0), \ \ \  P(\infty ) = 1 , \ \ \
V(\infty ) = 1
\eqno (4.4)
$$
The problem now has only two
free parameters: $\cos{\theta_W}$ and $\beta$.
This may be seen by rescaling fields and coordinates in the operator
$\bar{O}$ as in (4.1). With these rescalings the quantities $S_+$ and $P_+$ in
the eigenvalue problem (3.36) get replaced by: $$ S_+^* = {P^2\over 2} - cos^2
\theta_W {{{V'}^2} \over {P^*_+}}
       - {R \over 2} {d \over {dR}} \left \{
          {{V'} \over R} {{(n - 2 cos^2 \theta_W V ) } \over {P^*_+}}
                                   \right \}
$$
where,
$$
P^*_+ = (1 - 2 cos^2 \theta _W V )^2 + 2 cos^2 \theta _W R^2 P^2
$$

The rescaled eigenvalue problem was
solved by using a fifth order Runge-Kutta algorithm.
We kept $\beta$ fixed and found $\theta_w$ for which the
lowest eigenvalue changes sign. We repeated this procedure for several
values of $\beta$ and found the corresponding values of critical parameters
$(\sqrt{\beta}, \sin^2 \theta_w)$.
The above method was used to scan the range $0.07 \leq \beta \leq 1.0$.
Lower values of $\beta$ make the numerical analysis
fairly intensive since then there are two widely different scales in the
problem corresponding to the two widely different masses.
Our results are shown in Fig. 1  where we
plot the critical values of $\sqrt{\beta}$ (the ratio of the Higgs mass
to the $Z$
mass) versus the corresponding values of $\sin^2 \theta_w$.  In sector III, on
the right-hand side of the data line,
equation (3.36) had no negative eigenvalues implying
string stability.
Thus we may distinguish three sectors in Fig. 1:
sector I where the electroweak strings are unstable, sector III where strings
are stable, and,
the presently unexplored region shown as sector II
($\beta < 0.07$ or $m_H < 24 GeV$).
It is evident that the
physically realized values: $\sin^2 \theta_w=0.23$ and
$\sqrt{\beta}=m_H / m_Z > 0.62$
(see Ref. 14) lie entirely inside sector I.
This brings us to the main result of this paper: if the standard electroweak
model is the physically realized model, then the existing
vortex solutions in the bare model are unstable.

\beginsection{5. Outlook}

We have rigorously established that the electroweak model admits stable
vortex solutions for a certain range of parameters. This result is
exciting in that it makes the possibility of observing coherent states
in particle physics closer to reality. On the other hand, it is somewhat
dissappointing that the values of the parameters that Nature has
actually chosen are such that the vortex is unstable. However, this
still does not mean that the vortex will be unstable in the real world
since our analysis only applies to the bare electroweak model. In the
context of the early universe, for example, we must do a stability
analysis at high temperatures. One should also consider the possibility
that Nature has chosen
an extension of the standard model where the Higgs potential is more
complicated. In such circumstances, the stability issue would have
to be readdressed.

\bigskip

\beginsection {\it{Acknowledgements:}}

We would like to thank Miguel Ortiz for his help with the numerical
calculations. This work was supported by a CfA postdoctoral fellowship
(L.P.) and by the NSF(T.V.). M.J. acknowledges a research studentship from
SERC.

\vfill
\eject
{\centerline {References}}
\parindent 0 pt

1. T. Vachaspati and A. Ach\'ucarro, Phys. Rev. D {\bf 44},
3067 (1991).

2. G. W. Gibbons, M. Ortiz, F. Ruiz-Ruiz and T. Samols,
DAMTP preprint (1992); J. Preskill, Cal Tech preprint (1992).

3. M. Hindmarsh, Phys. Rev. Lett. {\bf 68}, 1263 (1992).

4. A. Ach\'ucarro, K. Kuijken, L. Perivolaropoulos and
T. Vachaspati, Nucl. Phys. B, to be published.

5. T. Vachaspati, Phys. Rev. Lett. {\bf 68}, 1977 (1992).

6. T. Vachaspati, Tufts preprint (1992).

7. T. Vachaspati and M. Barriola, Tufts preprint (1992).

8. S. Weinberg, Phys. Rev. Lett. {\bf 19}, 1264 (1967);
A. Salam in ``Elementary Particle Theory'', ed. N. Svarthholm, Stockholm:
Almqvist, Forlag AB, pg 367.

9. We are only considering the stability under small
perturbations. Hence, the word ``stability'' should always be understood
as ``meta-stability''.

10. J. C. Taylor, ``Gauge Theories of Weak Interactions'',
Cambridge University Press, 1976.

11. H. B. Nielsen and P. Olesen, Nucl. Phys. B{\bf{61}}, 45 (1973).

12. ``Solitons and Particles'', ed. C. Rebbi and G. Soliani,
World Scientific, 1984.

13. Barred indices always range from 1 to 2.

14. J. Steinberger, Phys. Rep. {\bf 203}, 345 (1991).

15. M. James, L. Perivolaropoulos and T. Vachaspati, to appear in Phys. Rev.
{\bf D}
(rapid communication).

\vfill
\eject

\head{Figure Captions}

 A map of parameter space showing the results of the stability analysis.
Sector I contains unstable strings, sector III contains stable strings
and we have not explored sector II. We also indicate the physically
allowed range of parameter space. The data from LEP constrains the Higgs
mass to $M_H >53 GeV$ which implies $\sqrt{\beta} > 0.62 GeV$ and the observed
$\sin^2 \theta_W$ is 0.23.

\endjnl
\end